\documentclass[10pt, a4paper]{article}
\usepackage{supertabular, setspace,hyperref}
\usepackage{amsmath,amssymb,amsthm,textcomp}
\usepackage{amsfonts,graphicx}
\usepackage{epstopdf}
\usepackage[mathscr]{eucal}
\usepackage{color}
\usepackage{float}
\usepackage{subfigure}
\usepackage{tikz}
\usetikzlibrary
{shapes,arrows,chains,matrix,positioning,scopes,shadows,calc}
\theoremstyle{definition}

\numberwithin{equation}{section}

\newcommand{\ncom}{\newcommand}

\ncom{\beq}{\begin{equation}}
\ncom{\eeq}{\end{equation}}
\ncom{\bea}{\begin{eqnarray*}}
\ncom{\eea}{\end{eqnarray*}}
\ncom{\beqa}{\begin{eqnarray}}
\ncom{\eeqa}{\end{eqnarray}}
\ncom{\nno}{\nonumber}
\ncom{\non}{\nonumber}
\ncom{\ds}{\displaystyle}
\ncom{\half}{\frac{1}{2}}
\ncom{\mbx}{\makebox{.25cm}}
\ncom{\hs}{\mbox{\hspace{.25cm}}}
\ncom{\rar}{\rightarrow}
\ncom{\Rar}{\Rightarrow}
\ncom{\noin}{\noindent}
\ncom{\bc}{\begin{center}}
\ncom{\ec}{\end{center}}
\ncom{\sz}{\scriptsize}
\ncom{\rf}{\ref}
\ncom{\s}{\sqrt{2}}
\ncom{\sgm}{\sigma}
\ncom{\Sgm}{\Sigma}
\ncom{\psgm}{\sigma^{\prime}}
\ncom{\dt}{\delta}
\ncom{\Dt}{\Delta}
\ncom{\lmd}{\lambda}
\ncom{\Lmd}{\Lambda}
\ncom{\Th}{\Theta}
\ncom{\e}{\eta}
\ncom{\eps}{\epsilon}
\ncom{\pcc}{\stackrel{P}{>}}
\ncom{\lp}{\stackrel{L_{p}}{>}}
\ncom{\dist}{{\rm\,dist}}
\ncom{\sspan}{{\rm\,span}}
\ncom{\re}{{\rm Re\,}}
\ncom{\im}{{\rm Im\,}}
\ncom{\sgn}{{\rm sgn\,}}
\ncom{\ba}{\begin{array}}
\ncom{\ea}{\end{array}}
\ncom{\hone}{\mbox{\hspace{1em}}}
\ncom{\htwo}{\mbox{\hspace{2em}}}
\ncom{\hthree}{\mbox{\hspace{3em}}}
\ncom{\hfour}{\mbox{\hspace{4em}}}
\ncom{\vone}{\vskip 2ex}
\ncom{\vtwo}{\vskip 4ex}
\ncom{\vonee}{\vskip 1.5ex}
\ncom{\vthree}{\vskip 6ex}
\ncom{\vfour}{\vspace*{8ex}}
\ncom{\norm}{\|\;\;\|}
\ncom{\integ}[4]{\int_{#1}^{#2}\,{#3}\,d{#4}}
\ncom{\vspan}[1]{{{\rm\,span}\{ #1 \}}}
\ncom{\dm}[1]{ {\displaystyle{#1} } }
\ncom{\ri}[1]{{#1} \index{#1}}

\newtheoremstyle
   {remarkstyle}
   {}
   {11pt}
   {}
   {}
   {\bfseries}
   {:}
   {     }
   {\thmname{#1} \thmnumber{#2} }

\theoremstyle{remarkstyle}

\def\eps{\varepsilon}

\begin{document}
\title{\bf Dissipative Waves in Real Gases}
\author{\bf Neelam Gupta and V. D. Sharma\\
{\it Department of Mathematics,
Indian Institute of Technology Bombay,}\\
{\it Powai, Mumbai-400076}}
\date{}
\maketitle
\begin{abstract}
In this paper, we characterize a class of solutions to the unsteady 2-dimensional flow of a van der Waals fluid involving shock waves, and derive an asymptotic amplitude equation exhibiting quadratic and cubic nonlinearities including dissipation and diffraction. We exploit the theory of nonclassical symmetry reduction to obtain some exact solutions. Because of the nonlinearities present in the evolution equation, one expects that the wave profile will eventually encounter distortion and steepening which in the limit of vanishing dissipation culminates into a shock wave; and once shock is formed, it will propagate by separating the portions of the continuous region. Here we have shown how the real gas effects, which manifest themselves through the van der Waals parameters $\tilde{a}$ and $\tilde{b}$ influence the wave characteristics, namely the shape, strength, and decay behavior of shocks.	
\end{abstract}
\vone
\section{Introduction}
It is well known that shock waves appearing in a wide range of physical systems are compressive in the sense that these are formed from characteristics approaching the discontinuity from both sides; examples include shock waves in perfect gases and longitudinal waves in solids. Studies have shown that in certain physical systems such as fluids with high specific heats, where the sign of the parameter $\Gamma$- so called fundamental derivative, defined as
$\Gamma(\rho,\mbox{\scriptsize{S}})={(c{\rho}^{-1}+ c_{\rho})}\vert_{\mbox{\scriptsize{S}}}$,
dictates the existence of sonic expansion shocks. Here  $\rho$, $\mbox{\scriptsize{S}}$, and $c$ denote, respectively, the density, entropy, and the speed of sound (see \cite{Thompson, Cramer1, Dunn, Kluwick}).
Here, we examine weakly nonlinear dissipative waves in the unsteady two-dimensional Navier-stokes equations governing the flow of a van der Waals fluid in which the fundamental derivative of gasdynamics changes sign in the vicinity of $\Gamma=0$.
In case of a perfect gas, $\Gamma$ is always positive and is $O(1)$, but for a  van der Waals fluid, $\Gamma$  may change sign in the pressure density plane and is $O(\epsilon)$ in contrast with the perfect gas case; here $\epsilon$ is a measure of the wave amplitude with $0<\epsilon<<1$. Thus, in the neighborhood of $\Gamma=0$, for perceptible nonlinear effects, we need time scales longer by an order of magnitude, necessitating the use of fast variables of a higher order magnitude to describe the propagation of signals with perturbed strength $O(\epsilon)$. We then use the method of multiple scales to derive the evolution equation governing the propagation of finite amplitude, two-dimensional, weakly nonlinear dissipative waves in the spirit closer to \cite{Kluwick, hunter88} and use the method of nonclassical symmetries
\cite{Bruzon,Winter,Nucci, radha} to obtain some exact solutions. The method of nonclassical symmetries is an extension of Lie's classical method in the sense that it may yield more solutions than those obtained using the classical method. Using this approach, we have found some new exact solutions involving shocks.
\section{Formulation of the problem}
Consider the dissipative flow of a van der Waals fluid governed by the unsteady 2-dimensional Navier-Stokes equations
\begin{align}
&\rho_{t} + {u}{\rho}_x +{\rho}{u}_x +{v}{\rho}_y+{\rho}{v}_y ={0},\nonumber\\&  
{u}_{t} +u{u}_x + {v}{u}_y + \frac{c^2}{\rho}{\rho_x}+\frac{1}{C_v}(p/{\rho}+a \rho)S_x=\frac{\mu}{\rho}(\frac{1}{3}{v}_{yx}+\frac{4}{3}u_{xx}+u_{yy}),\nonumber\\&
{v}_{t} +u{v}_x + {v}{v}_y + \frac{c^2}{\rho}{\rho_y}+\frac{1}{C_v}(p/{\rho}+a \rho)S_y=\frac{\mu}{\rho}(\frac{1}{3}{u}_{yx}+\frac{4}{3}v_{yy}+v_{xx}),\nonumber\\&
\mbox{\scriptsize{S}}_{t}+{u}\mbox{\scriptsize{S}}_x+{v}\mbox{\scriptsize{S}}_y= \frac{\mu}{\rho T}\left({4}/{3}(u^2_x+v^2_y-{u_x}{v_y})+v^2_x+u^2_y+2{v_x}{u_y}\right)+\kappa/{\rho C_v}(\mbox{\scriptsize{S}}_{xx}+\mbox{\scriptsize{S}}_{yy})+\nonumber\\& 
~~\quad \quad \quad \quad \quad \quad \quad  \frac{\kappa}{\rho}\left(\frac{\gamma-1}{\rho(1-b\rho)}(\rho_{xx}+\rho_{yy})+\frac{1}{C^2_v}(\mbox{\scriptsize{S}}^2_x+\mbox{\scriptsize{S}}^2_y) +\frac{2(\gamma-1)}{C_v \rho(1-b \rho)}({\rho_x}{\mbox{\scriptsize{S}}_x}+{\rho_y}{\mbox{\scriptsize{S}}_y})  \right. \nonumber\\&
~\quad \quad \quad \quad \quad \quad \quad 
\left.+\frac{(\gamma-1)(\gamma-2+2b\rho)}{\rho^2(1-b\rho)^2}(\rho^2_x+\rho^2_y)\right), \label{equ1}
\end{align}
for variables $(\rho, u, v, \mbox{\scriptsize{S}})$, where $\rho$ is the density, $(u,v)$ are the velocity components, $T$ is the absolute temperature, $\mu$ is the viscosity, and $\kappa$ is the thermal conductivity.\\
Here, we consider the gas that obeys the van der Waals type equation of state characterized by
\begin{equation}\label{equ2}
p=R\frac{T\rho}{1-b\rho}-a{\rho}^2,~~~~~\mbox{\scriptsize{S}}=R \ln \left(KT^{1/(\gamma-1)}\frac{1-b\rho}{\rho}\right), 
\end{equation}
where $p$. $T$, $\mbox{\scriptsize{S}}$, and $R$ denote pressure, temperature, entropy, and the gas constant, respectively; parameters $a$ and $b$ represent a measure for the attraction between constituent particles and the effective volume of each particle, respectively; $K$ is a positive constant, and $\gamma$ is the ratio of specific heats with values lying in the interval $1 <\gamma\leq 5/3$.
Using (\ref{equ2}), one can obtain the expression for sound speed 
\begin{equation}\label{equ3}
c=\sqrt{(p_{\rho})_{\mbox{\scriptsize{S}}}}=\left(\frac{\gamma(p+a\rho^2)}{\rho(1-b\rho)}-2a\rho\right)^{1/2}.
\end{equation}
For $c$ to be real and positive, we require that $0\leq b{\rho}<1$ and $0 \leq a\rho^2/p<1$.
In view of (\ref{equ2}) and (\ref{equ3}), equations (\ref{equ1}) can be written in the following vector matrix notation\\
\begin{align}
\textbf{U}_t+ A^1(\textbf{U})\textbf{U}_x + A^2(\textbf{U})\textbf{U}_y =& L^1(\textbf{U})\textbf{U}_{xx}+L^2(\textbf{U})\textbf{U}_{xy}+L^3(\textbf{U})\textbf{U}_{yy}+K^1(\textbf{U})Q(\textbf{U}_x, \textbf{U}_x) \nonumber \\&
+ K^2(\textbf{U})Q(\textbf{U}_x, \textbf{U}_y) + K^3(\textbf{U})Q(\textbf{U}_y, \textbf{U}_y), \label{equ4}
\end{align}
where
\[ \textbf{U}=\left( \begin{array}{c}
\rho \\
u \\
v\\
\mbox{\scriptsize{S}} \end{array} \right),~~~~~~
A^1=\left( \begin{array}{cccc}
u & \rho & 0 & 0 \\
\frac{c^2}{\rho} & u & 0 & \frac{(p/{\rho}+a \rho)}{C_v} \\
0 & 0 & u & 0 \\
0 & 0 & 0 & u \end{array} \right),~~~~~~
A^2=\left( \begin{array}{cccc}
v & 0 & \rho & 0 \\
0 & v & 0 & 0 \\
\frac{c^2}{\rho} & 0 & v & \frac{(p/{\rho}+a \rho)}{C_v} \\
0 & 0 & 0 & v \end{array} \right),
\]
\[ L_1=\left( \begin{array}{cccc}
0 & 0 & 0 & 0 \\
0 &\frac{ 4 {\mu}}{3 \rho} & 0 & 0 \\
0 & 0 & \frac{\mu}{\rho} & 0\\
\frac{\tilde{\beta}}{C_v} & 0 & 0 & \frac{\kappa}{\rho C_v}\end{array} \right),~~
L_2=\left( \begin{array}{cccc}
0 & 0 & 0 & 0 \\
0 & 0 &  \frac{\mu}{3 \rho} & 0 \\
0 &  \frac{\mu}{3 \rho} & 0 & 0 \\
0 & 0 & 0 & 0 \end{array} \right),~~
L_3=\left( \begin{array}{cccc}
0 & 0 & 0 & 0 \\
0 & \frac{\mu}{\rho} & 0 & 0 \\
0 & 0 & \frac{4{\mu}}{3\rho} & 0 \\
\frac{\tilde{\beta}}{C_v} & 0 & 0 & \frac{\kappa}{\rho C_v} \end{array} \right),
\]
\[ K_1=\left( \begin{array}{cccc}
0 & 0 & 0 & 0 \\
0 & 0 & 0 & 0 \\
\tilde{\alpha} & \frac{4\mu}{3\rho T} & \frac{\mu}{\rho T} & \frac{\kappa}{\rho C^2_v} \end{array} \right),~~~
K_2=\left( \begin{array}{cccc}
0 & 0 & 0 & 0 \\
0 & 0 &  0 & 0 \\
0 &  0 & 0 & 0 \\
\tilde{\beta} & -\frac{4\mu}{3\rho T} & \frac{2\mu}{\rho T} & \tilde{\beta} \end{array} \right),~~~
K_3=\left( \begin{array}{cccc}
0 & 0 & 0 & 0 \\
0 & 0 & 0 & 0 \\
0 & 0 & 0 & 0\\
\tilde{\alpha} & \frac{\mu}{\rho T} & \frac{4\mu}{3\rho T} & \frac{\kappa}{\rho C^2_v} \end{array} \right),\]
\[ Q(\textbf{U}_x, \textbf{U}_x)=\left( \begin{array}{c}
\rho^2_x \\
u^2_x \\
v^2_x\\
\mbox{\scriptsize{S}}^2_x \end{array} \right),~~~~~~
Q(\textbf{U}_x, \textbf{U}_y)=\left( \begin{array}{c}
\rho_x\mbox{\scriptsize{S}}_x \\
u_x v_y \\
v_x u_y\\
\rho_y S_y \end{array} \right),~~~~~~
Q(\textbf{U}_y, \textbf{U}_y)=\left( \begin{array}{c}
\rho^2_y \\
u^2_y \\
v^2_y\\
\mbox{\scriptsize{S}}^2_y \end{array} \right),
\]
with $\tilde{\alpha}=\dfrac{\kappa(\gamma-1)(\gamma-2+2b\rho)}{\rho^3(1-b\rho)^2}$ and $\tilde{\beta}=\dfrac{\kappa(\gamma-1)}{C_v\rho^2(1-b\rho)}$.
\section{Derivation of evolution equations}
In this section, we derive transport equation describing the evolution of the signal taking into account the dissipative effects and the variations in direction transverse to the characteristic rays. As outlined in the introduction, we introduce the variables
\begin{equation}\label{equ5}
\theta=\frac{\phi(x,y,t)}{\epsilon^2},~~~~~~\eta=\frac{\psi(x,y,t)}{\epsilon},
\end{equation}
where $\phi$ is the phase function and $\eta$ describes the modulations of wavefront along and transverse to the characteristic rays that lie on the surface $\psi=0$.
We look for asymptotic solutions of (\ref{equ4}) as $\epsilon \rightarrow 0$  of the form:
\begin{equation}\label{equ6}
\textbf{U}= \textbf{U}_0 + {\epsilon}\textbf{U}_1(x, y, t, \theta, \eta) + {\epsilon^2}\textbf{U}_2(x, y, t, \theta, \eta) + {\epsilon^3}\textbf{U}_3(x, y, t, \theta, \eta) +o(\epsilon^3),
\end{equation}
where $\textbf{U}_0=(\rho_0, 0, 0, \mbox{\scriptsize{S}})$ is the background state and $\textbf{U}_i$'s are smooth bounded functions of their arguments.
Using Taylor series expansion, we expand the coefficient matrices $A^1(\textbf{U})$ and $A^2(\textbf{U})$ as 
\begin{equation}\label{equ7}
A^r_{ij}=A^r_{ij}({U}_0)+B^r_{ijk}({U}^k-{U}^k_0)+C^r_{ijkl}({U}^k-{U}^k_0)({U}^l-{U}^l_0)+\ldots,
\end{equation} 
where $B^r_{ijk}=\dfrac{\partial{A^r_{ij}}}{\partial{{U}^k}}(\textbf{U}_0)$ and $C^r_{ijkl}=\dfrac{\partial^2{A^r_{ij}}}{2\partial{{U}^k}\partial{U^l}}(\textbf{U}_0)$.\\
Moreover, we consider the following expansions for $B^r_{ijk}$ and $C^r_{ijkl}$:
\begin{equation}\label{equ8}
B^r_{ijk}=B^r_{0ijk}+{\epsilon}B^r_{1ijk}+o(\epsilon),~~~~~~C^r_{ijkl}=C^r_{0ijkl}+{\epsilon}C^r_{1ijkl}+o(\epsilon).
\end{equation}
We consider both viscosity and heat conductivity to be $O(\epsilon^4)$, i.e., $\mu=\epsilon^4{\hat{\mu}}$ and $\kappa=\epsilon^4\hat{\kappa}$. In view of (\ref{equ5})-(\ref{equ8}), equation (\ref{equ4}) yields at levels $\epsilon^0$, $\epsilon^1$, and $\epsilon^2$ the following system of PDEs
\begin{align}\label{equ9}
\begin{split}
&O(\epsilon^{0}):~(\underbrace{{\phi_t}\delta_{ij}+A^1_{0ij}\phi_x + A^2_{0ij}\phi_y}_{M_{ij}})U^j_{1\theta}=0,\\&
O(\epsilon^1):~~(\underbrace{{\psi_t}\delta_{ij}+A^1_{0ij}\psi_x + A^2_{0ij}\psi_y}_{N_{ij}})U^j_{1\eta}+M_{ij}U^j_{2\theta}+(B^1_{0ijk}\phi_x + B^2_{0ijk}\phi_y)U^k_1U^j_{1\theta}=0\\&
O(\epsilon^2):~~M_{ij}U^j_{3\theta}+B^1_{0ijk}U^k_1U^j_{2\theta}+(B^1_{0ijk}U^k_2+B^1_{1ijk}U^k_1+C^1_{0ijkl}U^k_1U^l_1)U^j_{1\theta}+N_{ij}U^j_{2\eta}\\&~~~~~~~~~~~+B^2_{0ijk}U^k_1U^j_{1\eta}+U^i_{1t}+A^1_{0ij}U^j_{1x}+A^2_{0ij}U^j_{1y}=(\hat{L_{01}})_{ij}U^j_{1\theta\theta}.
\end{split}
\end{align}
It follows from $(\ref{equ9})_1$ that for a non-trivial solution  $U^j_1$, $\phi$  satisfies conditions $\det{M_{ij}}=0$, implying thereby that
\begin{equation}\label{equ10}
\phi^2_t=c^2_0(\phi^2_x+\phi^2_y).
\end{equation}
Let $L$ and $R$ be the left and the right null vectors of $M_{ij}$. Then $(\ref{equ9})_1$ implies that $U^j_{1}$ must be collinear to $R^j$,i.e.,
\begin{equation}\label{equ12}
U^j_{1}=h(x,y,t,\theta,\eta)R^j,
\end{equation}
where $h$ is a scalar valued function representing the wave amplitude and $L$ and $R$ are given as
\begin{equation*}
L=\left(-\dfrac{\phi_t}{\rho_0}, ~\phi_x,~ \phi_y, 
-\dfrac{(p_0/{\rho_0}+a\rho_0)}{c_0C_v}\phi_t\right),~~~~~~~~~~~R=\left(-\dfrac{\rho_0}{c^2_0}\phi_t, ~\phi_x, ~\phi_y, ~0\right)^T.
\end{equation*}
Multiplying $(\ref{equ9})_2$ by the left null vector $L$ of $M_{ij}$ and using (\ref{equ10}) and (\ref{equ12}), the resultant equation yields the following compatible relation
\begin{equation}\label{equ13}
\phi_t\psi_t-c^2_0(\phi_x\psi_x+\phi_y\psi_y)=0,
\end{equation}
with $h_{\eta} \neq 0$ and $\phi_t \neq 0$. We choose $\phi$ and $\psi$ satisfying (\ref{equ10}) and (\ref{equ13}), i.e.,
\begin{equation}\label{equ11}
\phi=x-c_0t, ~~~~~\psi=y.
\end{equation}
Let $S^j$ and $T^j$ be the vectors such that
\begin{equation}\label{equ15}
M_{ij}S^j=N_{ij}R^j~~~\text{ and}~~~~ M_{ij}T^j=B^1_{0ij}R^jR^k.
\end{equation}
Then the second equation in the set (\ref{equ9}) can be written as
\begin{equation*}
M_{ij}\left(U^j_2+S_j \int^{\theta}h_{\eta}+T^j(h^2/2)\right)=0,
\end{equation*}
which, in turn, implies that $\left(U^j_2+S_j \int^{\theta}h_{\eta}+T^j(h^2/2)\right)_{\theta}$ is collinear to the right null vector of $M_{ij}$, implying thereby that  $(\ref{equ9})_2$ has a self-consistent solution of the form 
\begin{equation}\label{equ14}
U^j_2=-gS^j-{(h^2/2)}T^j+\nu R^j,
\end{equation}
where $\nu$ is a arbitrary function of $\eta$ and 
\begin{equation} \label{equ02}
g_{\theta}=h_{\eta}.
\end{equation}
Premultiplying $(\ref{equ9})_3$ by $L$ and using (\ref{equ12}) and (\ref{equ14}), we find that the wave amplitude $h$ satisfies
\begin{equation}\label{equ16}
h_{\tau}-{\hat{\Gamma}_0}hh_{\theta}-{\Lambda_0}h^2h_{\theta}+(c_0/2)g_{\eta}=\hat{\beta}h_{\theta\theta},
\end{equation}
where $\tau$ is a temporal variable defined as $\partial_{\tau}=\partial_t+c_0\partial_x$, genuinely nonlinear parameter $\hat{\Gamma}_0$ that accounts for the nonlinear self interaction of waves is defined by 
\begin{equation}\label{equ17}
\hat{\Gamma}_0= \Gamma_0/{\epsilon} =\dfrac{ -\gamma (\gamma+1)(p_0+a\rho^2_0)+6a\rho^2_0(1-b\rho_0)^2}{2\epsilon (1-b\rho_0)\left(\gamma(p_0+a\rho^2_0)-2a\rho^2_0(1-b\rho_0)\right)}=O(1),
\end{equation}
parameter $\hat{\beta}$ which accounts for the dissipative effects is given by
\begin{equation}\label{equ17**}
\hat{\beta}=\frac{2\hat{\mu}}{3\rho_0}+\frac{\hat{\kappa}(\gamma-1)(p_0/\rho_0+a\rho_0)}{2C_v c^2_0\rho_0(1-b\rho_0)},
\end{equation}
and $\Lambda_0$, characterizing the real gas effects, is given by
\begin{equation}\label{equ18}
\Lambda_0=\dfrac{\sqrt{\rho_0}(\gamma(\gamma+1)(4-\gamma-6b\rho_0)(p_0+a\rho^2_0)-12a\rho^2_0(1-b\rho_0)^3)}{2(\gamma(p_0+a\rho^2_0)(1-b\rho_0)-2a\rho^2_0(1-b\rho_0)^2)^{3/2}}=O(1).
\end{equation}
Here, we restrict our discussion to the values of $\gamma$ and the dimensionless parameters $\tilde{a}$ and $\tilde{b}$, defined as $\tilde{a}=a(p_0/{\rho^2_0})^{-1}$ and $\tilde{b}=b\rho_0$, lying in the intervals $1<\gamma\leq 5/3$, $0\leq\tilde{a}<1$, and $0\leq\tilde{b}<1/3$ for which $\Lambda_0$ remains positive; however, $\hat{\Gamma}_0$ can take both positive and negative values.
The evolution of wave is governed by coupled system of transport equations (\ref{equ02}) and (\ref{equ16}), which we have been seeking. It may be observed that (\ref{equ16}) contains both quadratic and cubic nonlinearities inherent in the governing system (\ref{equ4}). The presence of the cubic nonlinearity is due the fact that higher order terms, neglected in the derivation of evolution equation, are of the same order as the quadratic nonlinear term; consequences of cubic nonlinearity, within the context of hyperbolic systems, are discussed in \cite{Kluwick}.
For smooth solutions, equations (\ref{equ02}) and (\ref{equ16}) can be combined to yield a single transport equation for the wave amplitude $h$
\begin{equation}\label{equ19*}
\left(h_{\tau}-{\hat{\Gamma}_0}hh_{\theta}-{\Lambda_0}h^2h_{\theta}-\hat{\beta}h_{\theta\theta}\right)_{\theta}+(c_0/2)h_{\eta \eta}=0,
\end{equation}
describing the propagation of weakly nonlinear and weakly diffracting two dimensional waves in a van der Waals gas; here $\tau$ is a temporal variable defined earlier, $\theta$ is a spatial variable in the propagation direction, and $\eta$ is the transverse spatial variable. It is apparent from (\ref{equ19*}) that the underlying structure in which the hyperbolic system is  embedded involves both dissipative and diffracted effects, and therefore we expect that the presence of quadratic and cubic nonlinearities cause steepening which would eventually lead to multivalued solution, but this development is opposed by the dissipative parameter $\hat{\beta}$. Thus, we expect shocks in a thin region of the profile.
\section{Nonclassical symmetries and group invariant solutions}
In order to find special solutions for (\ref{equ19*}), we perform the transformation
\begin{equation}\label{equ20*}
h({\tau},\theta,\eta)=H(\widetilde{T},\widetilde{X}),~~~~g({\tau}, \theta, \eta)=G(\widetilde{T},\widetilde{X}),
\end{equation}
with
\begin{equation}\label{equ21*}
\widetilde{X}=\theta-q{\tau},~~~{\widetilde{T}}=\eta,
\end{equation}
representing a wave structure propagating with velocity $q$. In view of (\ref{equ20*}) and (\ref{equ21*}), equations (\ref{equ02}) and (\ref{equ16}) become
\begin{equation}\label{equ032}
\left(-qH-\frac{{\hat{\Gamma}_0}}{2}H^2-\frac{{\Lambda_0}}{3}H^3\right)_{\widetilde{X}}+\frac{c_0}{2}G_{{\widetilde{T}}}=\hat{\beta}H_{\widetilde{X}\widetilde{X}},~~~~~~~~G_{\widetilde{X}}=H_{\widetilde{T}}.
\end{equation}
Equations (\ref{equ032}) imply that for smooth solutions, the wave amplitude $H$ satisfies the following equation
\begin{equation}\label{equ22*}
\left(\hat{\beta}H_{\widetilde{X}}+\frac{{\hat{\Gamma}_0}}{2}H^2+\frac{{\Lambda_0}}{3}H^3+qH\right)_{\widetilde{X}\widetilde{X}}-\frac{c_0}{2}H_{{\widetilde{T}}{\widetilde{T}}}=0.
\end{equation}
Since there is no general theory for finding the analytical solution of this type of PDE, symmetry group techniques provide a tool for obtaining special class of solutions \cite{Sergey}. Here, we use nonclassical method of symmetry analysis to reduce equation (\ref{equ22*}) into corresponding ODE and finally obtain solutions by analyzing the reduced equations.\\
Let us consider a one parameter Lie group of infinitesimal transformations
\begin{align}\label{equ24*}
\begin{split}
&\widetilde{X}^{*}=\widetilde{X}+{\delta}\xi(\widetilde{X},{\widetilde{T}},H)+O(\epsilon^2),\\&
{\widetilde{T}}^{*}={\widetilde{T}}+{\delta}\zeta(\widetilde{X},{\widetilde{T}},H)+O(\epsilon^2),\\&
H^{*}=H+{\delta}\chi(\widetilde{X},{\widetilde{T}},H)+O(\epsilon^2),
\end{split}
\end{align}
with a small parameter $\delta<<1$ and infinitesimals $\xi$, $\zeta$, and $\chi$. The vector field associated with the above group of transformations can be written as
\begin{equation}\label{equ23*}
V=\xi(\widetilde{X},{\widetilde{T}},H)\partial_{\widetilde{X}}+\zeta(\widetilde{X},{\widetilde{T}},H)\partial_{\widetilde{T}}+\chi(\widetilde{X},{\widetilde{T}},H)\partial_H
\end{equation}
The basic idea of the method is to require that equation (\ref{equ22*}) with invariance surface condition 
\begin{equation}\label{equ25*}
{\xi}H_{\widetilde{X}}+{\zeta}H_{\widetilde{T}}-\chi=0,
\end{equation}
remains invariant under the transformation (\ref{equ24*}) with infinitesimal operator (\ref{equ23*}) for infinitesimals $\xi$, $\zeta$, and $\chi$. It may be noticed that if $V$ is a nonclassical symmetry and $\lambda=\lambda(\widetilde{X},{\widetilde{T}},H)$ is an arbitrary function, then prolongation formula implies that $\lambda V$ is also a nonclassical symmetry. This property permits the normalization of any nonvanishing coefficient of the operator $V$ by setting it equal to one \cite{Sergey}.  Here, we consider two cases:\\
\textbf{Case 1: When} {$\mathbf{\zeta\neq 0}$}\\
Without loss of generality, we take $\zeta=1$ and then from the invariant surface condition (\ref{equ25*}), we obtain
\begin{align}\label{equ13*}
\begin{split}
&H_{\widetilde{T}}=\chi-\xi H_{\widetilde{X}},\\&
H_{{\widetilde{T}}{\widetilde{T}}}=(\chi_{\widetilde{T}}+\chi\chi_H-\xi\chi_{\widetilde{X}})-H_{\widetilde{X}}(2\xi\chi_H+\xi_{\widetilde{T}}+\chi\xi_H-\xi\xi_{\widetilde{X}})+2\xi\xi_H H^2_{\widetilde{X}}+\xi^2H_{\widetilde{X}\widetilde{X}}.
\end{split}
\end{align}
Eliminating $H_{{\widetilde{T}}{\widetilde{T}}}$ from equation (\ref{equ22*}), using $(\ref{equ13*})_2$, yields
\begin{align}
&\hat{\beta}H_{\widetilde{X}\widetilde{X}\widetilde{X}}-\left({{-\hat{\Gamma}_0}}H-{{\Lambda_0}}H^2-q+(c_0/2)\xi^2\right)H_{\widetilde{X}\widetilde{X}}-\left(-{{\hat{\Gamma}_0}}-{2{\Lambda_0}}H+c_0\xi\xi_H\right)H^2_{\widetilde{X}}  \nonumber\\&
+(c_0/2)(2\xi\chi_H+\xi_{\widetilde{T}}+\chi\xi_H-\xi\xi_{\widetilde{X}})H_{\widetilde{X}}-(c_0/2)(\chi_{\widetilde{T}}+\chi\chi_H-\xi\chi_{\widetilde{X}})=0.	 \label{equ14*}
\end{align}
Applying the fact that (\ref{equ22*}) remains invariant under the transformations (\ref{equ24*}), and eliminating $H_{\widetilde{X}\widetilde{X}\widetilde{X}}$ from the resulting equation with the help of (\ref{equ14*}), we get the following determining equations
\begin{equation}\label{equ016}
\xi=f(\widetilde{X},{\widetilde{T}}),~~~~~~\chi=\alpha(\widetilde{X},{\widetilde{T}})H+\beta(\widetilde{X},{\widetilde{T}}),
\end{equation}
where $f$, $\alpha$, and $\beta$ satisfy the following relations
\begin{align}
&3\hat{\beta}f_{\widetilde{X}\widetilde{X}}-3\hat{\beta}\alpha_{\widetilde{X}}-q\alpha+2qf_{\widetilde{X}}+(-\alpha+3f_{\widetilde{X}})(c_0f^2/2-q)+c_0\alpha f^2/2+c_0ff_{\widetilde{T}}-\hat{\Gamma}_0\beta=0, \nonumber \\&
2\Lambda_0\beta+\hat{\Gamma}_0\alpha+\hat{\Gamma}_0 f_{\widetilde{X}}=0, ~~~~
f_{\widetilde{X}}+2\alpha=0, ~~~~
\hat{\Gamma}_0f_{\widetilde{X}\widetilde{X}}-4\Lambda_0\beta_{\widetilde{X}}- 4\hat{\Gamma}_0\alpha_{\widetilde{X}}=0. \label{equ015}
\end{align}
Use of $(\ref{equ015})_3$ into $(\ref{equ015})_2$ yields $\beta=\hat{\Gamma}_0\alpha/{2\Lambda_0}$, which together with $(\ref{equ015})_4$ implies that
\begin{equation}\label{equ017}
\alpha_{\widetilde{X}}=0, ~~~\text{i.e.}, ~~~\alpha=f_1({\widetilde{T}}).
\end{equation}
Equation (\ref{equ017}), together with $(\ref{equ015})_3$ and $(\ref{equ016})_2$, yields the following expressions for $f$ and $\chi$
\begin{equation}\label{equ018}
f=-2f_1({\widetilde{T}})\widetilde{X}+f_2({\widetilde{T}})~~\text{and}~~\chi=f_1({\widetilde{T}})(H+\hat{\Gamma}_0/{2\Lambda_0}),
\end{equation}
where $f_1({\widetilde{T}})$ and $f_2({\widetilde{T}})$ satisfy the ODEs
\begin{align}\label{equ019}
\begin{split}
&f_{1{\widetilde{T}}{\widetilde{T}}}-4f_1f_{1{\widetilde{T}}}-6f^3_1=0,~~~~~~~~f_{1{\widetilde{T}}{\widetilde{T}}}+2f_1f_{1{\widetilde{T}}}-24f^3_1=0,\\&
f_{2{\widetilde{T}}{\widetilde{T}}}+2f_{1{\widetilde{T}}}f_2-24f^2_1f_2=0,~~~~f_2f_{2{\widetilde{T}}}-3f_1f^2_2+(2q/{c_0}-\hat{\Gamma}^2_0/{2c_0\Lambda_0})f_1=0.
\end{split}
\end{align}
The compatibility of $(\ref{equ019})_1$ and $(\ref{equ019})_2$ yields
\begin{equation}\label{equ020}
f_1(3f^2_1-f_{1{\widetilde{T}}})=0 \implies f_1({\widetilde{T}})=0,~ -1/{3({\widetilde{T}}+k_1)},
\end{equation}
where $k_1$ is an integration constant.
After using the values of $f_1$ into $(\ref{equ019})_3$ and $(\ref{equ019})_4$, we obtain the following pairs of functions
\begin{align}\label{equ021}
\begin{split}
&f_1({\widetilde{T}})=-1/{3({\widetilde{T}}+k_1)},~~~f_2({\widetilde{T}})=k_2/({\widetilde{T}}+k_1)~~~ \text{for} ~~~ 2q-\hat{\Gamma}^2_0/{2\Lambda_0}=0,\\&
f_1({\widetilde{T}})=0,~~~f_2({\widetilde{T}})=k_2,
\end{split}
\end{align}
where $k_2$  is an arbitrary constant.
Using (\ref{equ018}) and (\ref{equ021}) into (\ref{equ016}), we obtain from the corresponding pairs of functions the following expressions for the infinitesimals $\xi$ and $\chi$
\begin{equation}\label{equ022}
\xi=\frac{(2/3)\widetilde{X}+k_2}{{\widetilde{T}}+k_1},~~\chi=\frac{-1}{3({\widetilde{T}}+k_1)}(H+\hat{\Gamma}_0/{2\Lambda_0});~~~~~  ~~\xi=k_2,~~\chi=0.
\end{equation}
In view of (\ref{equ022}) and (\ref{equ23*}), the corresponding symmetry generators can be written as
\begin{equation}\label{equ023}
V=\frac{(2/3)\widetilde{X}+k_2}{{\widetilde{T}}+k_1}\partial_{\widetilde{X}}+\partial_{\widetilde{T}}+\frac{-1}{3({\widetilde{T}}+k_1)}(H+\hat{\Gamma}_0/{2\Lambda_0})\partial_H;~~~~~~~V=k_2\partial_{\widetilde{X}}+\partial_{\widetilde{T}}.
\end{equation}
In order to get exact solutions to equation (\ref{equ22*}), we use symmetry operators obtained in (\ref{equ023}); they reduce (\ref{equ22*}) into the corresponding ODEs.
From symmetry generator $(\ref{equ023})_1$, equation (\ref{equ22*}) possesses the following similarity reduction
\begin{equation}\label{equ025}
H=-\frac{\hat{\Gamma}_0}{2\Lambda_0}+\frac{F(\omega)}{({\widetilde{T}}+k_1)^{1/3}},~~~~~~~~~\omega=\frac{({\widetilde{X}}+({2}/{3})k_2)}{({\widetilde{T}}+k_1)^{2/3}},
\end{equation}
and $F(\omega)$ satisfies
\begin{equation}\label{equ026}
-\hat{\beta}F'''-{\Lambda_0}\left(F^2F''+2FF'^2\right)+{c_0}\left(\frac{2}{9}F+\frac{7}{9}F'\omega+\frac{2}{9}F''\omega^2\right)=0.
\end{equation}
In the limit of vanishing dissipation, i.e., $\hat{\beta}\rightarrow 0$, equation (\ref{equ026}) becomes
\begin{equation}\label{equ027}
\frac{\Lambda_0}{c_0}\left(F^2F''+2FF'^2\right)-\left(\frac{2}{9}F+\frac{7}{9}F'\omega+\frac{2}{9}F''\omega^2\right)=0,
\end{equation}
which admits 
\begin{equation}\label{equ010}
F(\omega)=\pm\sqrt{\frac{c_0}{2\Lambda_0}} \omega,
\end{equation} 
as particular solutions.
Equation $(\ref{equ025})_2$, on using (\ref{equ010}), yields
\begin{equation}\label{equ028}
H=-\frac{\hat{\Gamma}_0}{2\Lambda_0}\pm\sqrt{\frac{c_0}{2\Lambda_0}}\left(\frac{{\widetilde{X}}+(3/2)k_2}{{\widetilde{T}}+k_1}\right).
\end{equation}
The similarity solutions (\ref{equ028}) represent wavefans; it may be observed that due to invariance under translation in variables ${\widetilde{X}}$ and ${\widetilde{T}}$, the solution that we obtain will also be a solution by substituting ${\widetilde{X}}$ for ${\widetilde{X}}+(3/2)k_2$ and ${\widetilde{T}}$ by for ${\widetilde{T}}+k_1$; indeed the solutions (\ref{equ028}) blow up on the line ${\widetilde{T}}=-k_1$. In view of (\ref{equ028}), the solutions of (\ref{equ19*}) can be written as
\begin{equation}\label{equ028*}
h( {\tau}, \theta, \eta)=-\frac{\hat{\Gamma}_0}{2\Lambda_0}\pm\sqrt{\frac{c_0}{2\Lambda_0}}\left(\frac{\left(\theta-\frac{\hat{\Gamma}^2_0}{4\Lambda_0}{\tau}\right)+(3/2)k_2}{\eta+k_1}\right).
\end{equation}
Similarly, from symmetry generator $(\ref{equ023})_2$, the solution of (\ref{equ19*}) is given by\\
\begin{equation}\label{equ030}
h(\tau, \theta, \eta)=-\hat{\Gamma}_0/{2\Lambda_0}.
\end{equation}
\textbf{Case 2: When} $\mathbf{\zeta= 0}$ \textbf{and} $\mathbf{\xi\neq 0}$ \\
Without loss of generality, we take $\xi=1$. In this case the invariant surface condition (\ref{equ25*}) yields 
\begin{align}\label{equ07}
\begin{split}	
&H_{\widetilde{X}}=\chi({\widetilde{X}},{\widetilde{T}},H), ~~~~~H_{{\widetilde{X}}{\widetilde{X}}}=\chi_{\widetilde{X}}+\chi_H H_{\widetilde{X}},\\&
H_{{\widetilde{X}}{\widetilde{X}}{\widetilde{X}}}=\chi_{{\widetilde{X}}{\widetilde{X}}}+2\chi \chi_{{\widetilde{X}}{\widetilde{T}}}+\chi^2\chi_{HH}+\chi_{H}\chi_{{\widetilde{X}}}+\chi\chi^2_{H}.
\end{split}
\end{align}
Eliminating $H_{\widetilde{X}}$, $H_{{\widetilde{X}}{\widetilde{X}}}$, and $H_{{\widetilde{X}}{\widetilde{X}}{\widetilde{X}}}$ from equation (\ref{equ22*}) with the help of (\ref{equ07}), we obtain
\begin{align}\label{equ08}
\begin{split}
&\hat{\beta}(\chi_{{\widetilde{X}}{\widetilde{X}}}+2\chi\chi_{{\widetilde{X}}{\widetilde{T}}}+\chi^2\chi_{HH}+\chi_{H}\chi_{{\widetilde{X}}}+\chi\chi^2_{H})+\hat{\Gamma}_0(H(\chi_{\widetilde{X}}+\chi\chi_H)+\chi^2)\\&
+\Lambda_0(H^2(\chi_{\widetilde{X}}+\chi \chi_H)+2H\chi^2)+q(\chi_{\widetilde{X}}+\chi
\chi_H)-(c_0/2)H_{{\widetilde{T}}{\widetilde{T}}}=0
\end{split}
\end{align}
Applying the fact that (\ref{equ22*}) remains invariant under the transformations (\ref{equ24*}), and eliminating $H_{\widetilde{T}\widetilde{T}}$ from the resulting equation with the help of (\ref{equ08}), we get the following determining equations
\begin{equation}\label{equ26*}
\chi_{HH}=0 \implies \chi=\alpha({\widetilde{X}},{\widetilde{T}})H+\beta({\widetilde{X}},{\widetilde{T}}),
\end{equation}
where $\alpha$ and $\beta$ must satisfy the following nonlinear PDE system
\begin{align}
&\alpha_{\widetilde{T}}=0,~~~~~
\alpha_{{\widetilde{X}}{\widetilde{X}}}+8\alpha\alpha_{\widetilde{X}}+6\alpha^3=0,\nonumber\\&
\alpha_{{\widetilde{X}}{\widetilde{X}}}+5\alpha\alpha_{\widetilde{X}}+2\alpha^3+({\Lambda_0}/{\hat{\Gamma}_0})(\beta_{{\widetilde{X}}{\widetilde{X}}}+8\alpha_{\widetilde{X}}\beta+6\alpha\beta_{\widetilde{X}}+14\alpha^2\beta)=0,\nonumber\\&
\hat{\beta}(\alpha_{{\widetilde{X}}{\widetilde{X}}{\widetilde{X}}}+3\alpha\alpha_{{\widetilde{X}}{\widetilde{X}}}+3\alpha^2_{\widetilde{X}}+3\alpha^2\alpha_{\widetilde{X}})+2\Lambda_0(3\beta\beta_{\widetilde{X}}+5\alpha\beta^2)\nonumber\\&
+\hat{\Gamma}_0(\beta_{{\widetilde{X}}{\widetilde{X}}}+3\alpha\beta_{\widetilde{X}}+5\beta\alpha_{\widetilde{X}}+4\alpha^2\beta)
+(q\alpha_{{\widetilde{X}}{\widetilde{X}}}+2q\alpha\alpha_{\widetilde{X}}-(c_0/2)\alpha_{{\widetilde{T}}{\widetilde{T}}})=0,\nonumber\\&
\hat{\beta}(\beta_{{\widetilde{X}}{\widetilde{X}}{\widetilde{X}}}+3\beta\alpha_{{\widetilde{X}}{\widetilde{X}}}+3\beta\alpha\alpha_{\widetilde{X}}+3\alpha_{\widetilde{X}}\beta_{\widetilde{X}})+(q\beta_{{\widetilde{X}}{\widetilde{X}}}-(c_0/2)\beta_{{\widetilde{T}}{\widetilde{T}}}+2q\beta\alpha_{\widetilde{X}})
+\hat{\Gamma}_0(3\beta\beta_{\widetilde{X}}+2\alpha\beta^2)\nonumber\\& +{2\beta^3\Lambda_0}=0.\label{equ27*}
\end{align}
Equation $(\ref{equ27*})_2$ along with $(\ref{equ27*})_1$ yields 
\begin{equation}\label{equ0}
\alpha=f({\widetilde{X}}),~~~~\text{with}~~f= 1/3{\widetilde{X}},~~1/{\widetilde{X}},~\text{and}~0.
\end{equation}
For $f=1/{3{\widetilde{X}}}$, functions $\alpha({\widetilde{X}},{\widetilde{T}})=1/3{\widetilde{X}}$ and  $\beta({\widetilde{X}},{\widetilde{T}})=\dfrac{\hat{\Gamma}_0}{6\Lambda_0{\widetilde{X}}}$ satisfy the system (\ref{equ27*}) only if
\begin{equation}\label{equ05*}
\hat{\beta}=0 ~~~~\text{and}~~~~2q-\frac{\hat{\Gamma}^2_0}{2\Lambda_0}=0.
\end{equation}
Therefore, in this case the symmetry operator can be written as
\begin{equation}\label{equ05}
V=\partial_{\widetilde{X}}+(1/3{\widetilde{X}})(H+{\hat{\Gamma}_0}/{2\Lambda_0})\partial_H.
\end{equation}
and the similarity variables corresponding to (\ref{equ05}) are
\begin{equation}\label{equ29**}
\omega={\widetilde{T}},~~~H=-\hat{\Gamma}_0/{2\Lambda_0}+{\widetilde{X}}^{1/3}F(\omega),
\end{equation}
where  $F$, in view of (\ref{equ22*}), satisfies the ODE $F''=0$, the solution of which can be written as $F(\omega)=\bar{k}_1\omega+\bar{k}_2$, 
where $\bar{k}_1$ and $\bar{k}_2$ are integration constants. Thus, solution of equation (\ref{equ22*}) is given by
\begin{equation}\label{equ34*}
H=-\hat{\Gamma}_0/{2\Lambda_0}+{\widetilde{X}}^{1/3}(\bar{k}_1{\widetilde{T}}+\bar{k}_2),
\end{equation}
and hence the solution of (\ref{equ19*}) can be written as
\begin{equation}\label{equ34**}
h(\tau,\theta, \eta )= -\hat{\Gamma}_0/{2\Lambda_0}+\bar{k}_1\eta\left(\theta-\frac{\hat{\Gamma}^2_0}{4\Lambda_0}\tau\right)^{1/3}.
\end{equation} 
Following the same methodology, we  derive the solutions of (\ref{equ22*}) for $f({\widetilde{X}})=1/{\widetilde{X}}$ and $f({\widetilde{X}})=0$ as  
\begin{equation*}\label{equ33*}
H=-\frac{\hat{\Gamma}_0}{2\Lambda_0}\pm \sqrt{\frac{c_0}{2\Lambda_0}}\frac{{\widetilde{X}}}{{\widetilde{T}}}~~~\text{and}~~~H=-\frac{\hat{\Gamma}_0}{2\Lambda_0}\pm\sqrt{\frac{c_0}{2\Lambda_0}}\frac{{\widetilde{X}}}{{\widetilde{T}}}+\frac{\bar{k}_1}{{\widetilde{T}}}+\bar{k}_2{\widetilde{T}}^2,~~\text{respectively}.
\end{equation*}
\section{Solutions involving shocks}
As observed earlier, the solution of (\ref{equ22*}) may involve shocks.
Let ${\widetilde{X}}={\widetilde{X}}({\widetilde{T}})$ be the shock propagating into the medium where $H=H^{+}$(constant) is the state ahead of the shock while $H=H^{-}$ is the value of $H$ behind the shock.
Then the R-H conditions for the system (\ref{equ032}) can be written as
\begin{equation}\label{equ51*}
\left[qH+(\hat{\Gamma_0}/{2})H^2+({\Lambda_0}/{3})H^3\right]+(c_0/2)\left[G\right]\frac{d{\widetilde{X}}}{d{\widetilde{T}}}=0,~~~~~~[G]+[H]\frac{d{\widetilde{X}}}{d{\widetilde{T}}}=0,
\end{equation}
where $[H]=H^{-}-H^{+}$.
Elimination of $[G]$ from the set of equations (\ref{equ51*}) and on using (\ref{equ05*}), we get
\begin{equation}\label{equ36*}
\hat{\Gamma}_0^2/{4\Lambda_0}+(\hat{\Gamma_0}/{2})(H^{+}+H^{-})+({\Lambda_0}/{3})({H^{+}}^2+{H^{-}}^2+H^{+}H^{-})-\frac{c_0}{2}\left(\frac{d{\widetilde{X}}}{d{\widetilde{T}}}\right)^2=0,
\end{equation}
For values of $H$ behind the shock given by (\ref{equ028}) and ahead of the shock, where  $H=H^{+}=-\hat{\Gamma}_0/{2\Lambda_0}$, equation (\ref{equ36*}) yields $d{\widetilde{X}}/d{\widetilde{T}}=\pm {\widetilde{X}}/\sqrt{3}{\widetilde{T}}$, which on integration furnishes the shock trajectories
\begin{equation}\label{equ37*} {\widetilde{X}}=C_1{\widetilde{T}}^{1/\sqrt{3}},~~~~{\widetilde{X}}=C_2{\widetilde{T}}^{-1/\sqrt{3}},
\end{equation}
where $C_1$ and $C_2$ are the integration constants. The forward facing shock $(\ref{equ37*})_1$, which is followed by the wavefan represented by (\ref{equ028*}) with plus sign, assumes the following form in the $(\theta, \eta, \tau)$-space as
\begin{equation}\label{equ37**}
\theta=\frac{\hat{\Gamma}^2_0}{4\Lambda_0}\tau+C_1 \eta^{1/\sqrt{3}}, 
\end{equation}
followed by the wavefan represented by  (\ref{equ028*}) with plus sign, the strength of which is given by 
\begin{equation}\label{equ40}
[h]=\sqrt{\frac{c_0}{{2\Lambda_0}}}\eta^{-\alpha}, ~~~~~ \alpha = ({\sqrt{3}}-1)/{\sqrt{3}}
\end{equation}
that decays according to the law $[h] = O(\eta^{-\alpha})$ as $\eta \rightarrow \infty$. Moreover, the shock is compressive in the sense that $h$ decreases across the shock as we move from left to right. However, if the shock is followed by the state represented by (\ref{equ028*}) with minus sign, then the shock is expansive in the sense that $h$ increases across the shock as we move form left to right, and it decays according to the same law. A similar discussion follows for the backward facing shock propagating into a constant state and is followed by the wavefan represented by the exact solution (\ref{equ028*}) with plus or minus sign. 
\section{Results and conclusions}
\begin{figure}[h]
	\scalebox{1.2}{
		\subfigure[]{
			\includegraphics[width=1.8in,height=1.8in]{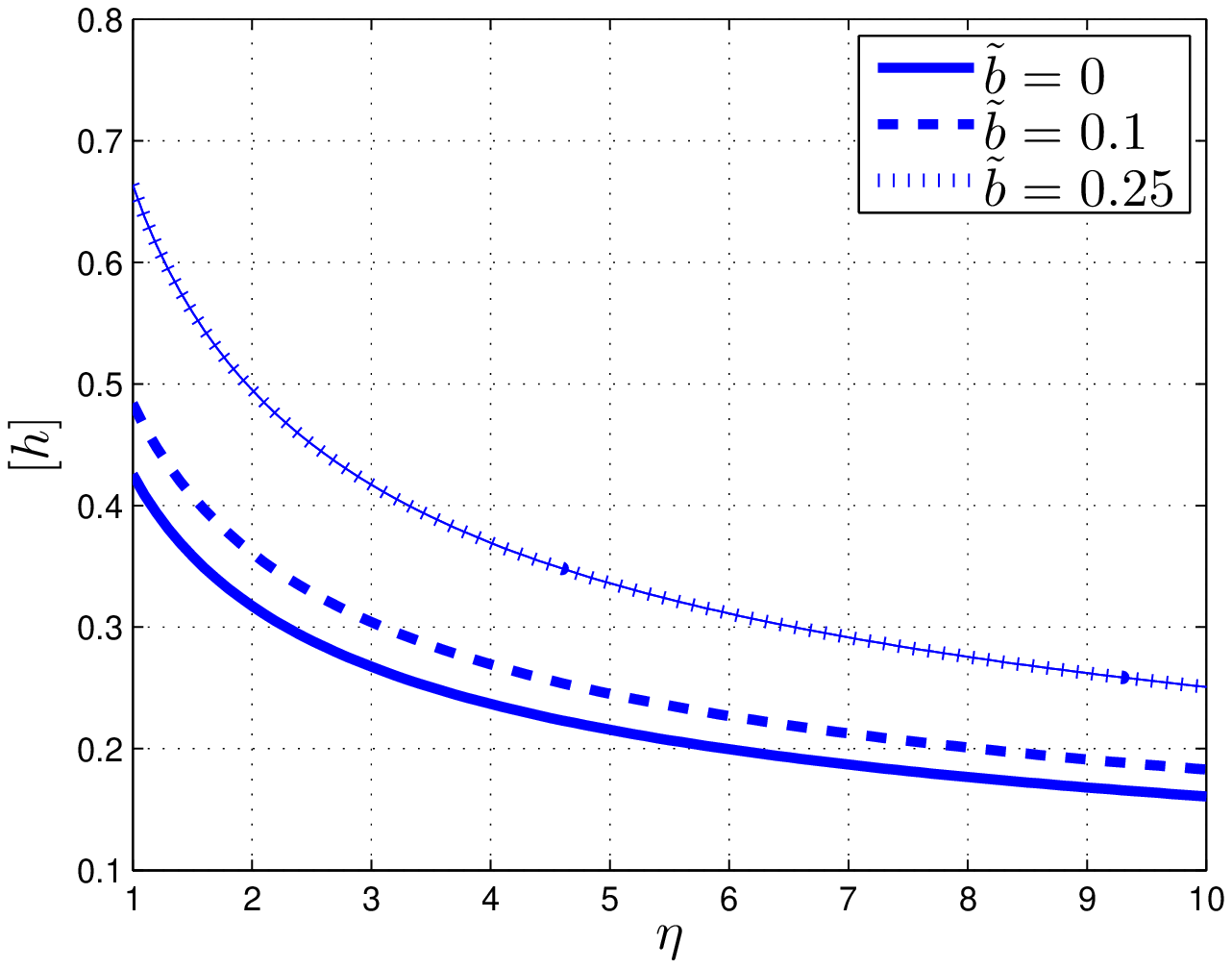}\label{figure4.1}
		}
		\hspace{-1cm}
		\subfigure[]{
			\includegraphics[width=2in,height=1.8in]{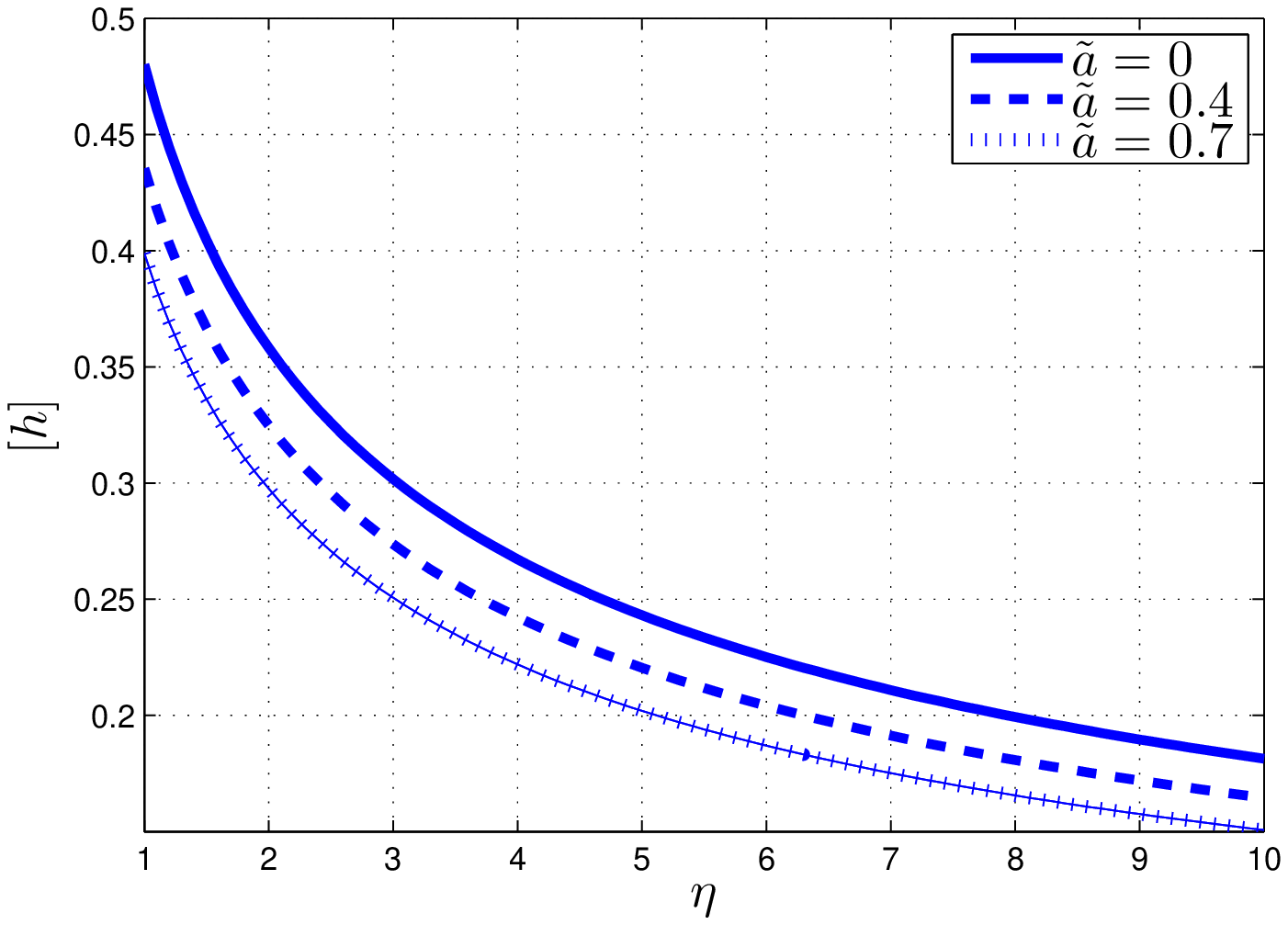}\label{figure4.2}
		}
		\hspace{-1cm}
		\subfigure[]{
			\includegraphics[width=2in,height=1.8in]{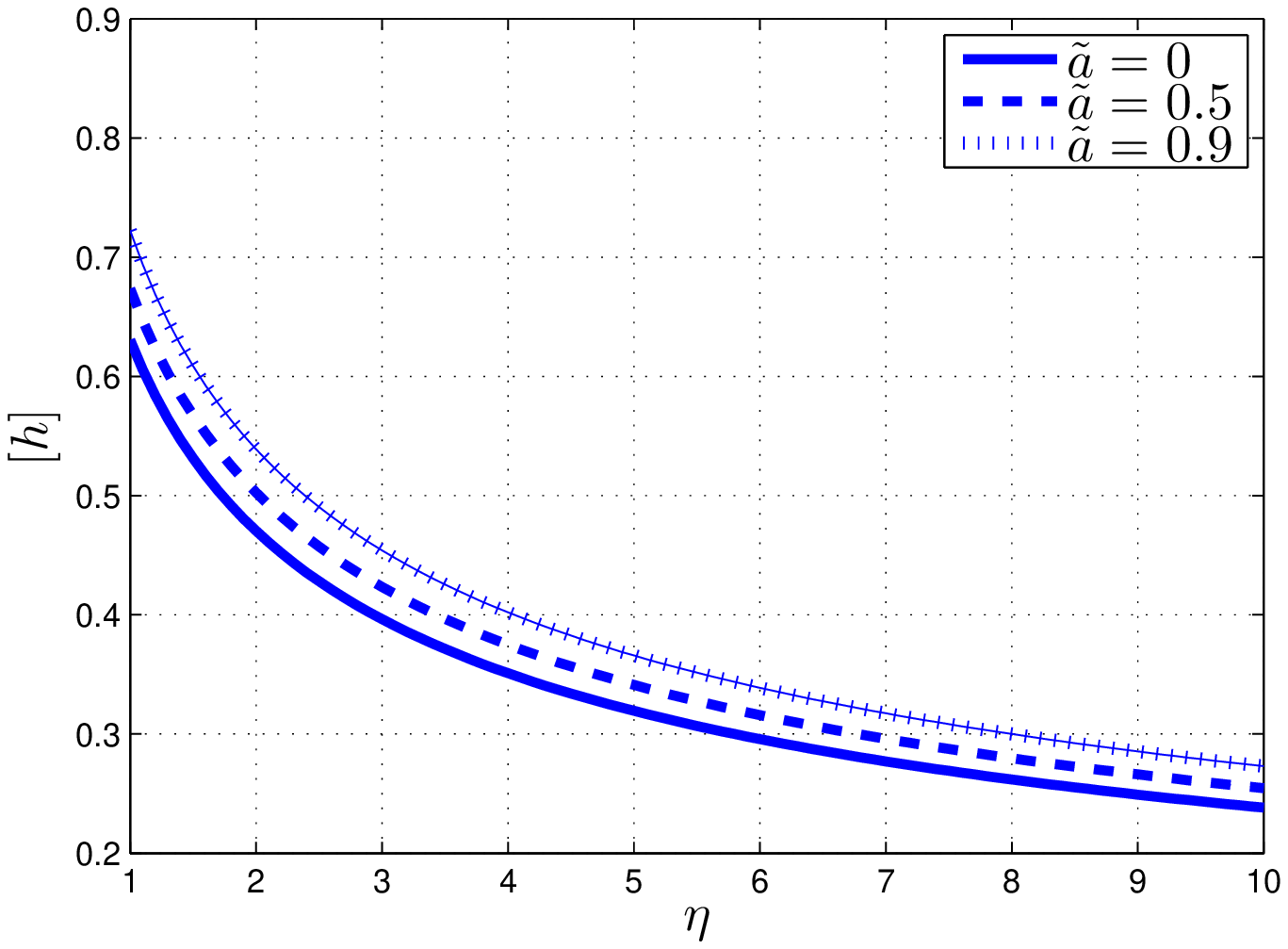}\label{figure4.3}
		}
	}
	\caption{\textit{(a) Shock strength and decay rate influenced by the parameter $\tilde{b}$  for fixed $\tilde{a}=0.4$ and $\gamma=1.4$, (b) Shock strength and decay rate influenced by the parameter $\tilde{a}$ for fixed $\gamma=1.4$ and $\tilde{b}=0.02<\tilde{b}^*$, and (c) Shock strength and decay rate influenced by the parameter $\tilde{a}$ for fixed $\gamma=1.4$ and $\tilde{b}=0.25>\tilde{b}^*$, respectively.}}
\end{figure}

\begin{figure}[h]
	\scalebox{1.5}{
		\subfigure[]{
			\includegraphics[width=1.8in,height=1.8in]{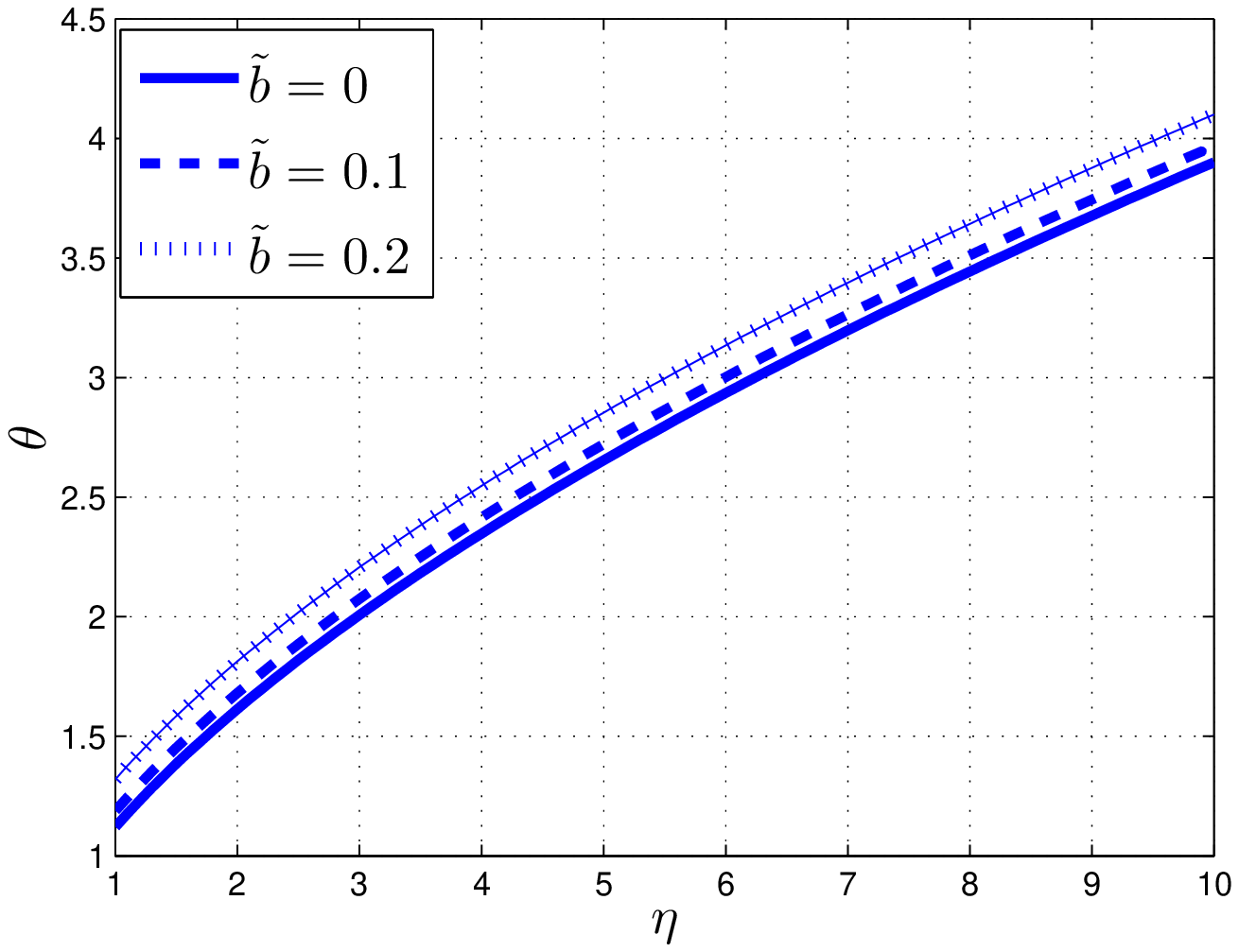}\label{figure4.4}
		}
		\hspace{0.5cm}
		\subfigure[]{
			\includegraphics[width=2in,height=1.8in]{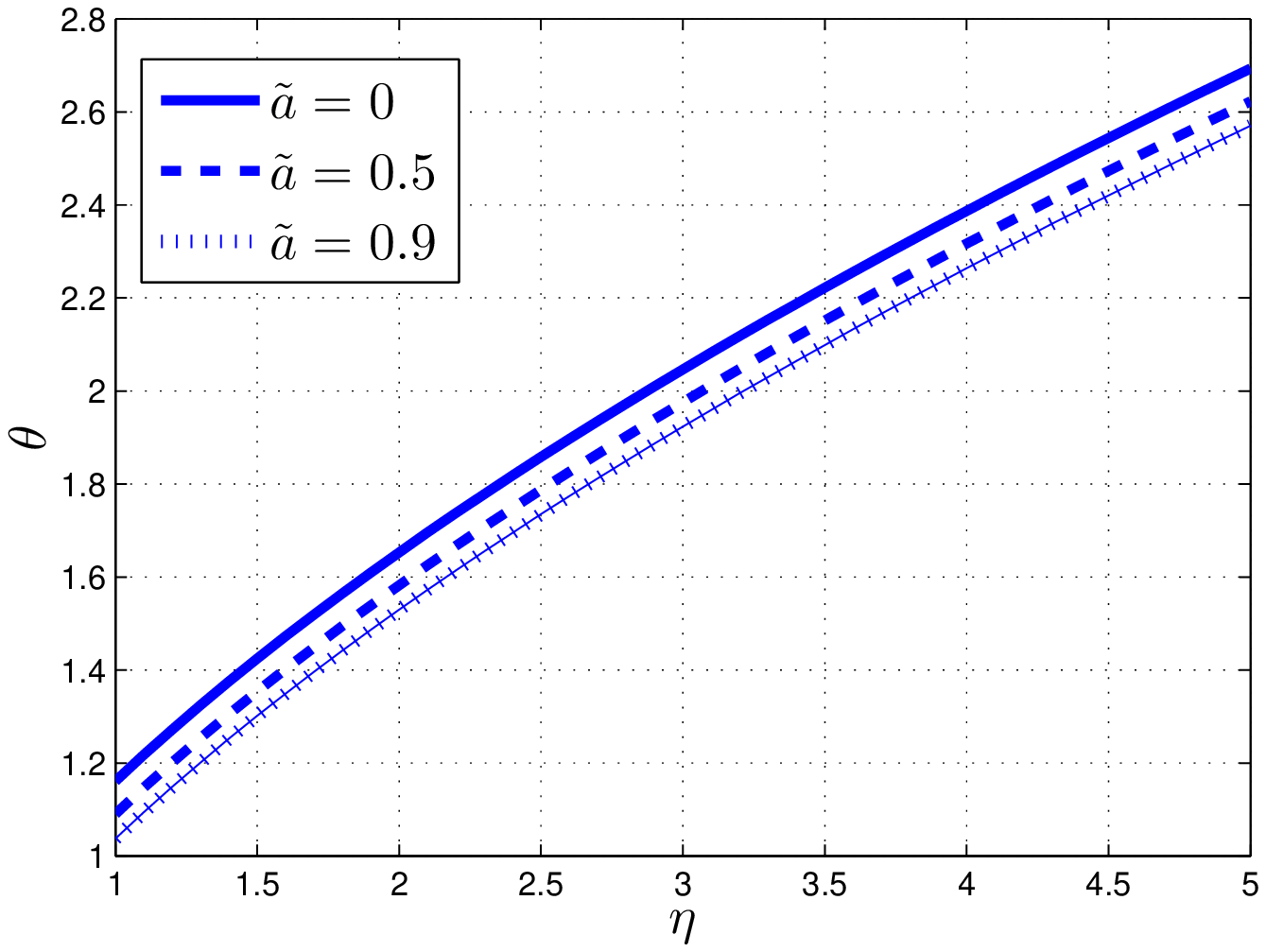}\label{figure4.5}
		}
	}
	\caption{\textit{(a) Shock trajectory influenced by the parameter $\tilde{b}$ for fixed $\gamma=1.4$ and $\tilde{a}=0.2$ and (b) Shock trajectory influenced by the parameter $\tilde{a}$ for fixed $\gamma=1.4$ and $\tilde{b}=0.02$.}}
\end{figure}

In this section, we explore how the real gas effects influence the shape, strength and decay law for the forward facing shock followed by a compressive wavefan represented by the solution (\ref{equ028*}) with plus sign. We find that an increase in the van der Waals parameter $\tilde{b}$ causes the shock strength to increase relative to what it would have been in the ideal gas case $(\tilde{a} = 0 = \tilde{b})$ (see Figure \ref{figure4.1}). However, for any value of  $\tilde{a}$ lying in the interval $0 \leq \tilde{a} < 1$, there exists a critical value $\tilde{b}^*$ of $\tilde{b}$, such that for values
of $\tilde{b}$ lying in the interval $(0, \tilde{b}^*)$ the strength of the compressive shock decreases (see Figure \ref{figure4.2}), whereas it increases for values of $\tilde{b}$ lying in the interval $(\tilde{b}^*,1/3)$; see Figure \ref{figure4.3}. This result is very much consistent with the conclusions arrived at in \cite{neelam}.
Moreover, the shape of the forward facing shock (see Figures \ref{figure4.4} and \ref{figure4.5}) as well as its decay rate (see Figures \ref{figure4.1}, \ref{figure4.2}, and \ref{figure4.3}) are also influenced by the parameters $\tilde{a}$ and $\tilde{b}$. An increase in the van der Waals parameter $\tilde{b}$ causes the  decay rate of the forward facing shock to decrease (see Figure \ref{figure4.1}); however, for a fixed $\tilde{a}$, there exists a critical value $\tilde{b}^*$ of $\tilde{b}$, such that for values of $\tilde{b}$ lying in the interval $(0, \tilde{b}^*)$ the shock decays faster (see Figure \ref{figure4.2}), whereas for values of $\tilde{b}$ lying in the interval $(\tilde{b}^*,1/3)$ the decay rate is slower (see Figure \ref{figure4.3}) than the ideal gas case. 

\end{document}